\def\lesssim{\mathrel{\hbox{\rlap{\hbox{\lower4pt\hbox{$\sim$}}}\hbox{$<$}}}}
\def\gtrsim{\mathrel{\hbox{\rlap{\hbox{\lower4pt\hbox{$\sim$}}}\hbox{$>$}}}}

\def\teff{$T_{\rm eff}$~}

\def\ll_lsun{log$({L/\rm L_{\odot}})$~}
\def\masa_msun{$M/ \rm M_{\odot}$~}
\def\m_mstar{$M/M_{*}$~}

\documentclass{aa}
\usepackage{graphicx}
        
\begin{document}

\title{Mass-radius relations for massive white dwarf stars}

\author{L. G. Althaus$^{1}$\thanks{Member of the Carrera del Investigador
        Cient\'{\i}fico y Tecnol\'ogico, CONICET, Argentina.},
        E. Garc\'{\i}a--Berro$^{1,2}$, 
        J. Isern$^{2,3}$ \and 
        A. H. C\'orsico$^{4,5\star}$}

\offprints{L. G. Althaus }

\institute{$^1$Departament   de  F\'\i   sica   Aplicada,  Universitat
           Polit\`ecnica de Catalunya, Av. del Canal Ol\'\i mpic, s/n,
           08860 Castelldefels, Spain\\   
           $^2$Institut d'Estudis  Espacials de Catalunya,  Ed. Nexus, 
           c/Gran Capit\`a 2, 08034 Barcelona, Spain.\\ 
           $^3$Institut  de  Ci\`encies  de  l'Espai, C.S.I.C., Campus 
           UAB, Facultat de  Ci\`encies, Torre  C-5, 08193 Bellaterra,
           Spain\\
           $^4$Facultad de Ciencias  Astron\'omicas y Geof\'{\i}sicas,
           Universidad Nacional de La Plata,  Paseo  del  Bosque  s/n,  
           (1900) La Plata, Argentina.\\  
           $^5$Instituto de Astrof\'{\i}sica La Plata, IALP, CONICET\\
\email{leandro@fa.upc.es,       garcia@fa.upc.es,   isern@ieec.fcr.es, 
       acorsico@fcaglp.unlp.edu.ar}}
\date{Received; accepted}

\abstract{We present  detailed  theoretical  mass-radius relations for 
massive  white  dwarf stars  with  oxygen-neon  cores.   This work  is
motivated  by recent  observational  evidence about  the existence  of
white dwarf stars  with very high surface gravities.   Our results are
based on evolutionary calculations that take into account the chemical
composition expected  from the  evolutionary history of  massive white
dwarf  progenitors. We present  theoretical mass-radius  relations for
stellar mass values ranging from 1.06 to $1.30\, M_{\sun}$ with a step
of  $0.02\, M_{\sun}$  and effective  temperatures from  150\,000~K to
$\approx$ 5\,000~K.   A novel aspect predicted by  our calculations is
that  the  mass-radius relation  for  the  most  massive white  dwarfs
exhibits a  marked dependence  on the neutrino  luminosity.  Extensive
tabulations for  massive white dwarfs,  accessible from our  web site,
are presented as well.
\keywords{dense matter  ---  stars:  evolution  ---  stars:  white  dwarfs }}

\authorrunning{Althaus et al.}

\titlerunning{Mass-radius relations for massive white dwarf stars}

\maketitle


\section{Introduction}

White  dwarfs  are  the  end-product  of the  evolution  of  low-  and
intermediate-mass stars. Thus, they preserve important clues about the
formation  and  evolution  of  our  Galaxy. This  information  can  be
retrieved by  studying their  observed mass, kinematic  and luminosity
distributions, provided that we  have good structural and evolutionary
models for  the progenitors of white  dwarfs and for  the white dwarfs
themselves.   In  particular, two  fundamental  tools  to analyze  the
properties  of the white  dwarf population  as a  whole are  the white
dwarf  luminosity   function  and  the  mass   distribution  of  white
dwarfs. The former  has been consistently used to  obtain estimates of
the age of  the Galactic disk (Winget et  al.  1987; Garc\'\i a--Berro
et al.  1988;  Hernanz et al. 1994; Richer et al.   2000) and the past
history of the galactic star  formation rate (Noh \& Scalo 1990; D\'\i
az--Pinto et al.  1994). The latter yields important information about
the late  stages of stellar evolution  since it reveals  the amount of
mass  lost  during stellar  evolution  from  an  initial stellar  mass
distribution  (Liebert,  Bergeron \&  Holberg  2005). Moreover,  white
dwarfs  have been  proposed to  contribute significantly  to  the mass
budget of  our Galaxy --- see  Garc\'\i a--Berro et al.   (2004) for a
recent discussion on this topic.

However,  the fundamental  parameters of  white dwarf  stars,  such as
their individual masses and  radii, are difficult to estimate.  Hence,
reliable  mass and  radius  determinations are  only  available for  a
relatively small  fraction of the whole local  white dwarf population.
Consequently, one  of the  most important theoretical  results, namely
the mass--radius relation, is  poorly constrained by the observations,
despite  the   improved  evolutionary  sequences   available  nowadays
(Benvenuto \& Althaus 1999; Hansen 1999; Salaris et al. 2000).

A  good determination of  the white  dwarf mass  distribution provides
useful  constraints to  the  initial--final mass  relationship, a  key
ingredient for  modelling the different white  dwarf populations (thin
and  thick disk,  spheroid\ldots), and  a touchstone  for  the stellar
evolutionary  models. Several works  have focused  on this  issue. For
instance, Finley  et al. (1997) determined  the effective temperatures
and   surface  gravities  of   174  DA   white  dwarfs   with  $T_{\rm
eff}>25\,000$~K, and from that  they obtained the mass distribution of
hot DA white  dwarfs.  Marsh et al. (1997) describe  the results for a
sample of  89 EUV-selected white  dwarfs and Vennes  (1999) determined
the  mass  distribution  for   a  sample  of  141  EUV-selected  white
dwarfs. Silvestri  et al.  (2001) studied  the gravitational redshifts
and  mass distribution  of 41  white  dwarfs in  common proper  motion
binary  systems. Finally,  Liebert et  al. (2005)  studied  a complete
sample of 348 DA white dwarfs from the Palomar Green survey. All these
studies report  the existence of a  clear, well defined  peak at $\sim
0.6\,M_{\sun}$  with  a typical  dispersion  of $\sim  0.1\,M_{\sun}$.
Additionally, a secondary peak is  also reported at masses between 1.0
and $1.2\,M_{\sun}$.   These high-mass objects should  have ONe cores,
the primary  products of the  $^{12}$C nuclear reactions,  in contrast
with average-mass  white dwarfs, which undoubtedly have  CO cores. The
origin of these  high-mass white dwarfs is uncertain,  and it has been
suggested that  they could  be the result  of either  binary evolution
(Marsh et al.  1997) --- but see  Guerrero et al. (2004) --- or of the
evolution of  heavy-weight intermediate-mass single  stars (Ritossa et
al.  1996; Garc\'\i  a--Berro et al.  1997; Iben  et al. 1997, Ritossa
et al.  1999).  In  any case these  white dwarfs  do exist and  a good
mass-radius relation is needed  in order to accurately determine their
respective masses.   All the  mass determinations, including  the most
recent ones, rely on the  evolutionary sequences of Wood (1995), which
are  valid for  CO cores.  To the  best of  our knowledge  no reliable
mass-radius relations  for massive white  dwarfs with ONe  cores exist
today and, thus, this is precisely  the goal of the present work.  Our
paper is  organized as follows. In  \S2 we present  our input physics.
In \S3  we discuss  our evolutionary sequences.  Finally, in  the last
section we summarize our findings and draw our conclusions.

\section{Input physics}

The evolutionary  code used in this  paper is that used  in our recent
work about  the evolution and pulsational properties  of massive white
dwarfs  with oxygen and  neon cores  (C\'orsico et  al. 2004),  and we
refer the reader to that paper for details about the input physics and
about  the  procedure  followed  to  obtain the  initial  white  dwarf
configurations.    The  code   is  based   on  updated   and  delailed
constitutive  physics.   Briefly,  the   equation  of  state  for  the
low-density  regime includes  an updated  version of  the  equation of
state  of Magni \&  Mazzitelli (1979).   For the  high-density regime,
partially    degenerate   electrons,    radiation    pressure,   ionic
contributions  and  Coulomb  interactions are  considered.   Radiative
opacities  are  those of  OPAL  (Iglesias  \&  Rogers 1996)  including
carbon- and oxygen-rich compositions, complemented at low temperatures
with the  Alexander \&  Ferguson (1994) molecular  opacities. Neutrino
emission rates  for pair, photo, plasma  and bremsstrahlung processes,
and  high-density conductive  opacities  are taken  from  Itoh et  al.
(1996a,b),  and Itoh  et al.   (1994).  Convection  is treated  in the
framework  of   the  mixing  length   theory  as  given  by   the  ML2
parameterization  (Tassoul et  al.   1990).  In  this  paper, we  have
neglected the effects of crystallization. Because we are interested in
the  calculation  of  mass  radius relations,  this  assumption  bears
virtually   no  relevance.    However,   it  is   worth  noting   that
crystallization  in massive  white  dwarfs takes  place  at such  high
stellar luminosities  that its  impact on the  white dwarf  cooling is
less  relevant.  In  addition,  chemical redistribution  due to  phase
separation  upon crystallizaton of  the oxygen-neon  core is  of minor
importance ---  see C\'orsico et al.   (2004) for details  --- and has
not been considered in this work.

\begin{figure}[t]
\centering
\includegraphics[clip,width=250pt]{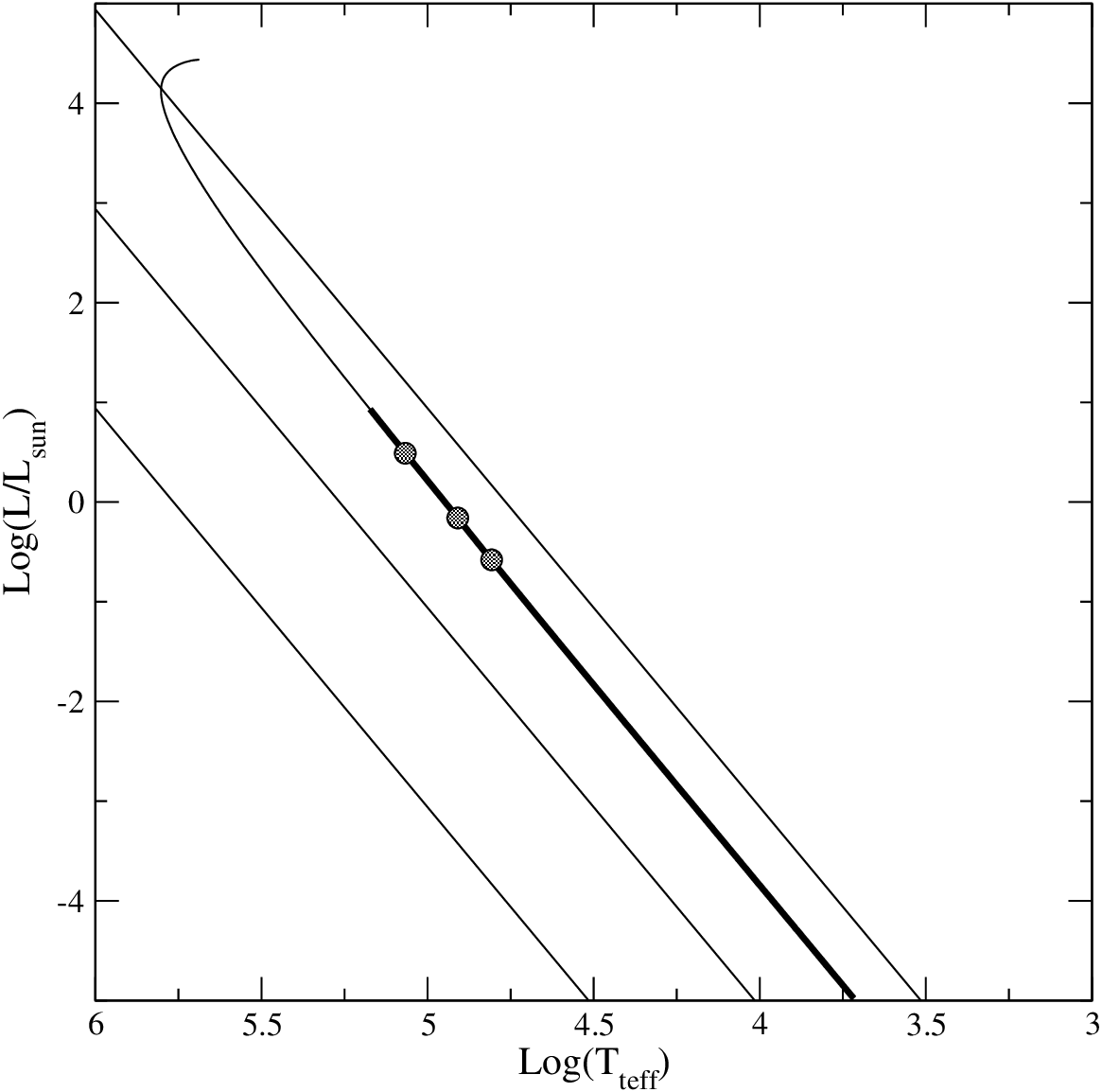}
\caption{Hertzsprung-Russell   diagram   for   two   $1.28\, M_{\sun}$ 
         evolutionary    sequences     with    different    artificial
         heating.  Lines of  constant radius  for $R=0.01$,  0.001 and
         $0.0001\, R_{\sun}$ (from top  to bottom) are also shown. See
         text for details.}
\label{fig01.eps}
\end{figure}

Our  starting stellar  configurations  correspond to  hot white  dwarf
structures  with a  realistic chemical  stratification  appropriate to
white dwarfs resulting from progenitor stars of solar metallicity that
are expected  to have burnt carbon in  semidegenerate conditions.  All
the  evolutionary sequences  considered  in this  work  have the  same
chemical  abundance profiles,  and correspond  to that  illustrated in
figure 4 of  C\'orsico et al.  (2004).  Although  minor changes in the
chemical  profile are  expected  because of  the  different masses  of
progenitor  objects,   these  have  a  negligible   influence  in  the
mass-radius  relation.    The  outer  layer   chemical  stratification
consists  of a pure  hydrogen envelope  of $1.4  \times 10^{-6}  M_* $
overlying a  pure helium shell of  $4 \times 10^{-4} M_*  $ and, below
that, a  buffer rich  in $^{12}$C and  $^{16}$O.  We mention  that the
amount of hydrogen we adopted is  an upper limit as imposed by nuclear
reactions.   We also find  that the  mass-radius relations  are almost
insensitive  to the  hydrogen envelope  mass.  The  core  is primarily
composed  of $^{16}$O  and $^{20}$Ne,  plus some  traces  of $^{12}$C,
$^{23}$Na and $^{24}$Mg.  The chemical  composition of the core is the
result of  repeated shell  flashes that take  place during  the carbon
burning phase in massive white dwarf progenitors (Garc\'{\i}a-Berro et
al.  1997).   The oxygen-neon core  of our models is  characterized by
flat  profiles  that are  the  result  of  a rehomogenization  process
induced  by  Rayleigh-Taylor instabilities  in  regions with  negative
molecular weight gradients,  a process that is expected  to take place
along  the white  dwarf evolutionary  sequence.  We  mention  that the
lowest  stellar mass  we  considered in  this  study is  close to  the
theoretical  lower  limit  of   $\approx  1.05  \,  M_{\sun}$  for  an
oxygen-neon white dwarf to be formed (Gil-Pons et al. 2003).

\begin{figure}[t]
\centering
\includegraphics[clip,width=250pt]{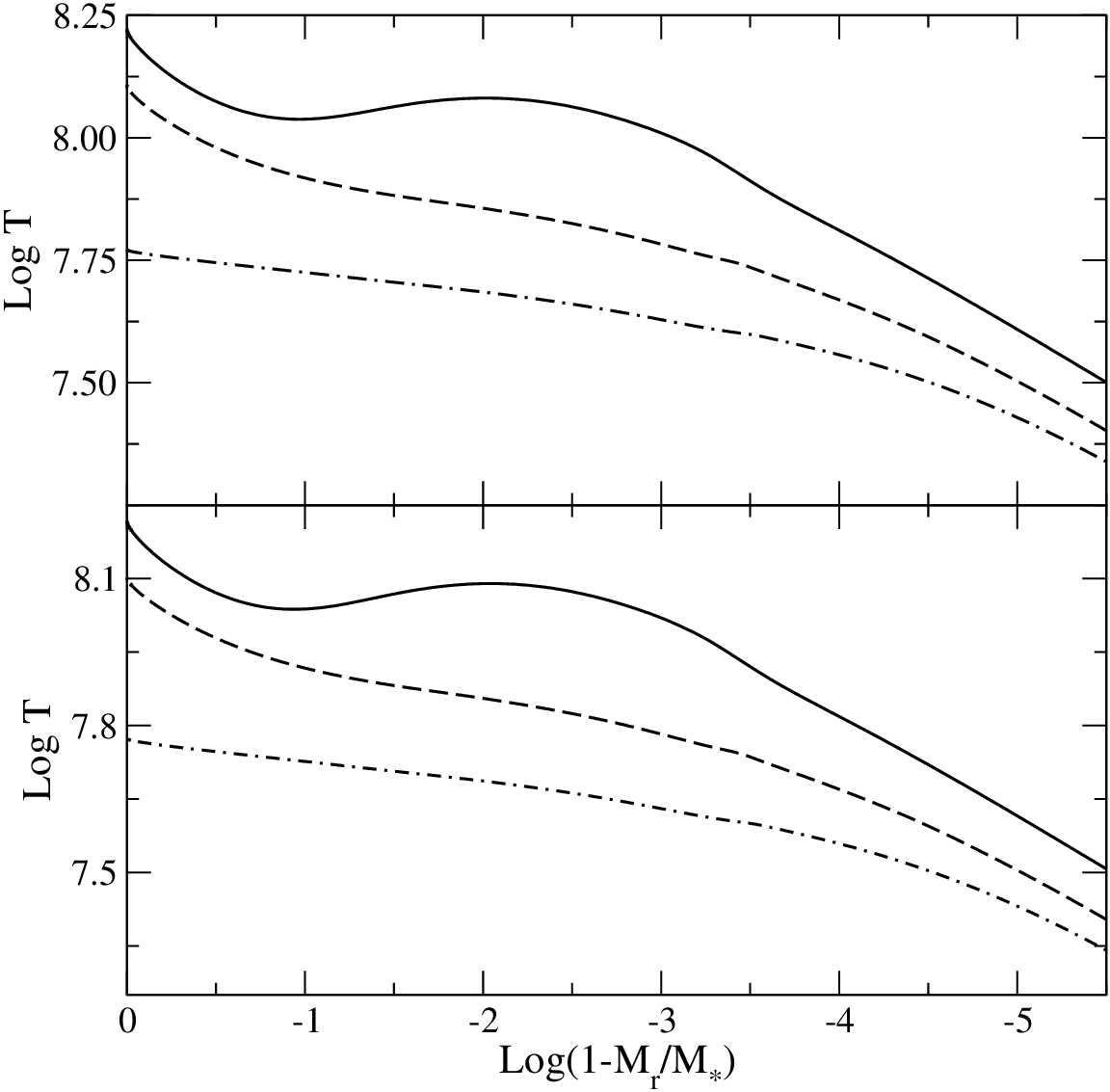}
\caption{Internal  temperature  in  terms of  the outer mass  fraction.
         The top  and bottom panel correspond,  respectively, to three
         selected models from the evolutionary sequences depicted with
         thin  and  thick  lines   in  figure  1.  Solid,  dashed  and
         dot-dashed  lines correspond to  luminosities of  about $\log
         L/L_{\sun}=$0.49,  $-0.16$  and  $-0.58$, respectively.  Note
         that  for   the  two  most  luminous   models,  the  internal
         temperature is strongly influenced by neutrino emission.}
\label{ltemp}
\end{figure}

The initial stellar models needed  to start our cooling sequences were
obtained  by  means  of  the same  artificial  evolutionary  procedure
described  in Gautschy  \&  Althaus (2002)  and  Althaus \&  C\'orsico
(2004).    Briefly, we  start from a  hot $0.95\,  M_{\sun}$ white
dwarf   model  obtained  {\sl   self-consistently}  by   evolving  the
progenitor star  initially on  the ZAMS (Althaus  et al. 2003)  to the
white  dwarf  regime.   In   order  to  obtain  initial  white  dwarf
configurations  for the different  masses studied  here we  scaled the
stellar  mass and  incorporated an  artificial energy  generation rate
until the white dwarf reaches a luminosity far larger than that of the
initial models  considered here as meaningful. We  checked the thermal
structure  of the  models  and  found that  the  transitory state  has
negligible influence  on the thermal configuration of  the white dwarf
interior well  before the initial  configuration is reached  (see next
section).   Specifically, we checked the thermal  structure of the
$1.28 \, M_{\sun}$ model by extending the heating artificial procedure
even   further  beyond   the  knee   at  high   luminosities   in  the
Hertzsprung-Russell diagram, as shown in Fig~\ref{fig01.eps}.  In this
figure, the thick line displays  the evolutionary track for one of our
standard sequences (that of the  $1.28 \, M_{\sun}$ model), whilst the
thin line  corresponds to the cooling track obtained  when the artificial
heating is extended  to much larger luminosities.  As  can be seen the
two  evolutionary tracks are  indistinguishable, thus  confirming that
the evolutionary results presented below  do not depend on the initial
conditions.

We have  computed the evolution of massive  white dwarf configurations
from hot  effective temperature stages down to  very low luminosities.
The values of  the stellar mass considered range  from 1.06 to $1.30\,
M_{\sun}$ with a  step of $0.02\, M_{\sun}$.  The  lower limit of this
mass range corresponds  to the minimum mass for a  white dwarf to have
an ONe core, whereas the upper limit corresponds to white dwarf masses
for   which  post-newtonian   corrections  start   to   be  important.
Nevertheless, it  is worth noting that extremely  massive white dwarfs
--- those  with $M_*  \ga 1.3\,  M_{\sun}$ ---  are very  unusual. The
mass-radius relations  are presented for  effective temperature values
ranging from about 150\,000~K to $\approx$ 5\,000~K.

\section{Evolutionary results}

\begin{figure}[t]
\centering
\includegraphics[clip,width=250pt]{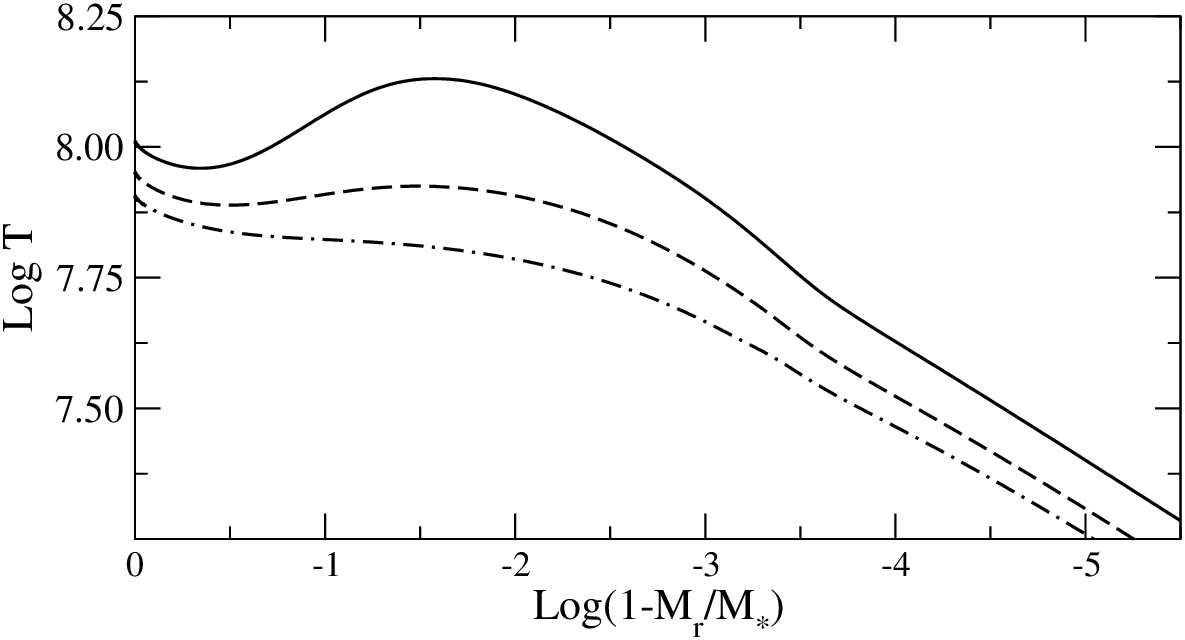}
\caption{Internal  temperature in  terms  of the  outer mass fraction.  
         Solid, dashed and dot-dashed  lines correspond to the $1.06\,
         M_{\sun}$  white  dwarf  models  at  $\log  L/L_{\sun}=$0.66,
         $-0.02$ and  $-0.38$, respectively.   Note that for  the most
         luminous model, the maximum temperature is off-center.}
\label{ltemp2}
\end{figure}

We  begin  by examining  Fig.~\ref{ltemp},  which  shows the  internal
temperature as a function of  the outer mass fraction, $\log (1-M_{\rm
r}/M_*)$,  for  three  selected  evolutionary stages  of  the  $1.28\,
M_{\sun}$ white  dwarf sequence corresponding, from top  to bottom, to
total  luminosities  of  about  $\log  L/L_{\sun}=0.49$,  $-0.16$  and
$-0.58$, respectively,   which are  marked as dots in  Fig.~1.  As
can  be seen  the three  models are  already at  advanced evolutionary
stages,   and,  consequently,  their   thermal  structure   should  be
reliable. The  top panel refers  to our standard sequence,  whilst the
bottom  panel  displays the  situation  for  models  belonging to  the
sequence that started at very  high luminosities (denoted by thin line
in Fig.~1). Note  that the temperature profiles of  the models are the
same for both sequences,  reflecting the correctness of our artificial
procedure  to generate  starting white  dwarf configurations.  

\begin{figure}[t]
\centering
\includegraphics[clip,width=250pt]{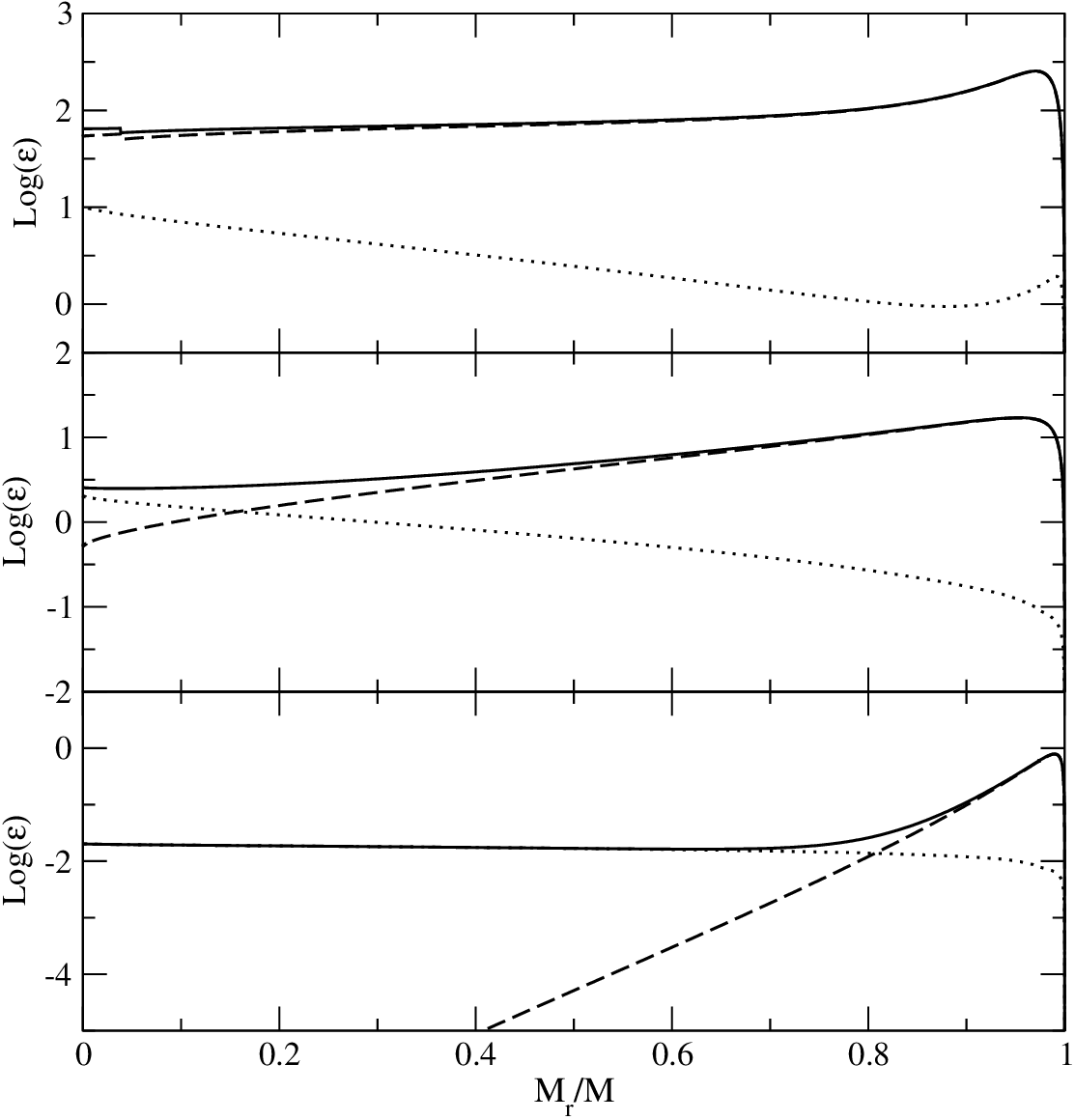}
\caption{Neutrino emission  rate (in erg/g/s) in  terms  of  the  mass 
         fraction for three  selected $1.28\,M_{\sun}$ stellar models.
         Solid, dotted  and dashed lines correspond to  the total rate
         of neutrino emission, and to the rates due to brehmsstrahlung
         and  plasma  processes, respectively.  The  top, central  and
         bottom  panel  correspond  to  luminosities  of  about  $\log
         L/L_{\sun}=$0.51, $-0.17$ and $-0.58$.}
\label{fig04.eps}
\end{figure}

It is  well known  that neutrino losses  considerably affect  both the
cooling time-scales  {\sl and} the structural properties  of hot white
dwarfs.   Indeed,  neutrino emission  constitutes  the most  important
physical process affecting the evolution  of very hot CO white dwarfs,
regardless of  their mass.   As can be  seen in  Figs.~\ref{ltemp} and
\ref{ltemp2}, this  statement is also  true for the massive  ONe white
dwarfs studied  here.  Note that, for  the most massive  model and for
the luminosity range considered here, neutrino losses do not cause the
maximum    temperature   to   occur    off-center   (top    panel   of
Fig.~\ref{ltemp}), as  it is  the case for  the less massive  model of
$1.06\,  M_{\sun}$,  which  is  displayed  in  Fig.~\ref{ltemp2}.   In
particular it  is worth  noting that for  the $1.28\,  M_{\sun}$ white
dwarf cooling  sequence, the model with the  largest photon luminosity
--- solid  lines in  Fig.~\ref{ltemp}  --- temperature  shows a  local
maximum  at $\log(1-M_{\rm  r}/M_*)\simeq -2.2$,  well inside  the ONe
core.   This can  be understood by examining Fig.~\ref{fig04.eps},
which shows the neutrino emission  rates (in erg/g/s) as a function of
the  mass  fraction  for  the  $1.28\,  M_{\sun}$  white  dwarf  model
sequence.  Note  that the  maximum neutrino emissivity  occurs towards
the outer edge of the  degenerate core, with plasma neutrino being the
dominant contribution. Indeed, for  massive white dwarfs, the emission
rate resulting from  plasma neutrino is characterized by  a maximum at
densities much  lower than  the central densities  of such  models, as
clearly shown, for instance, in Fig.~6 of Itoh et al. (1996b).

\begin{figure}[t]
\centering
\includegraphics[clip,width=250pt]{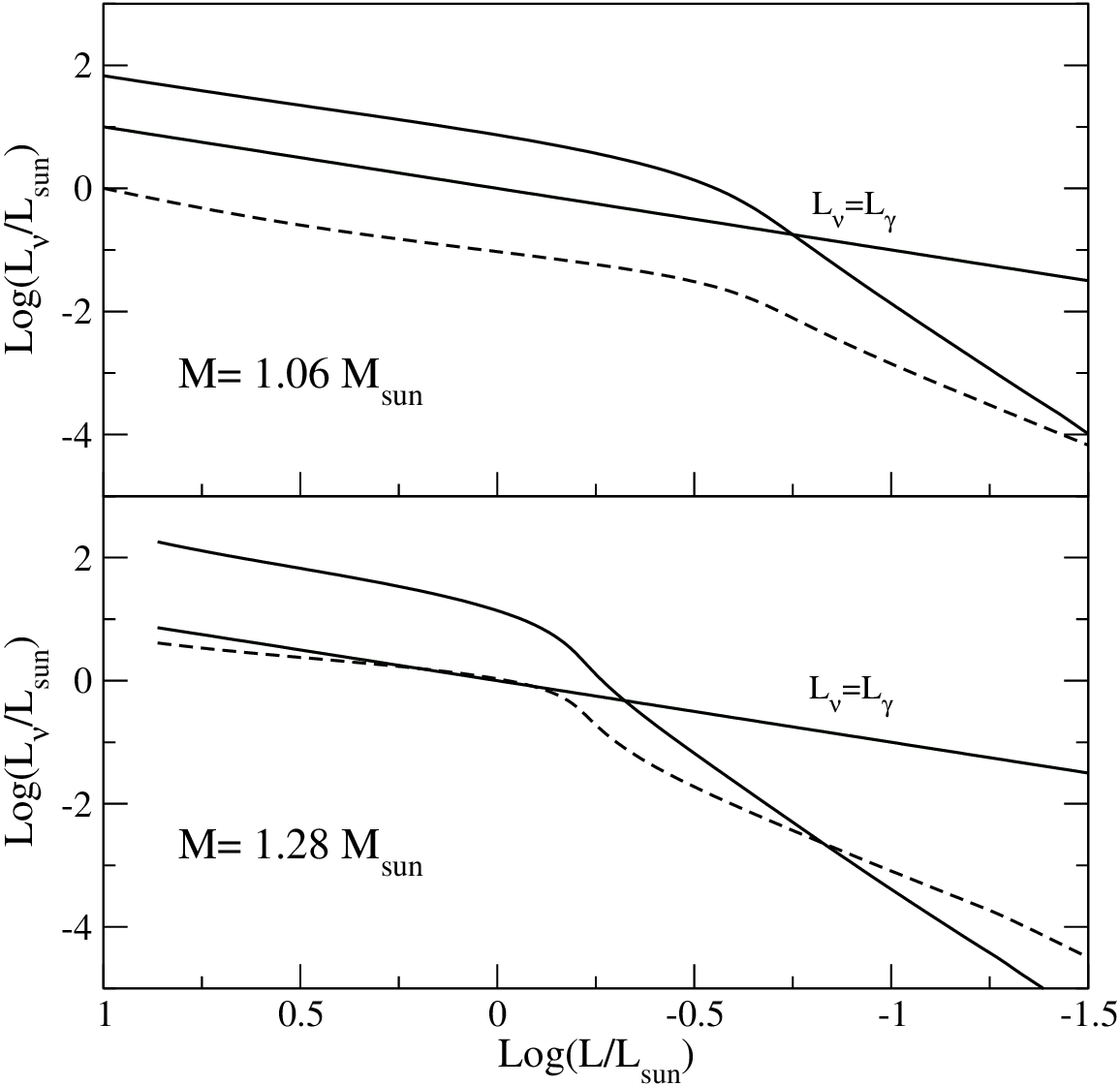}
\caption{Neutrino  emission   luminosity  resulting  from  plasma  and 
         bremsstrahlung    processes   (solid   and    dashed   lines,
         respectively) for the 1.06  and $1.28\, M_{\sun}$ white dwarf
         sequences.   In the  interests of  comparison, we  depict the
         situation for  which neutrino energy losses are  equal to the
         photon luminosity of the white dwarf.}
\label{procesos}
\end{figure}

\begin{figure}[t]
\centering
\includegraphics[clip,width=250pt]{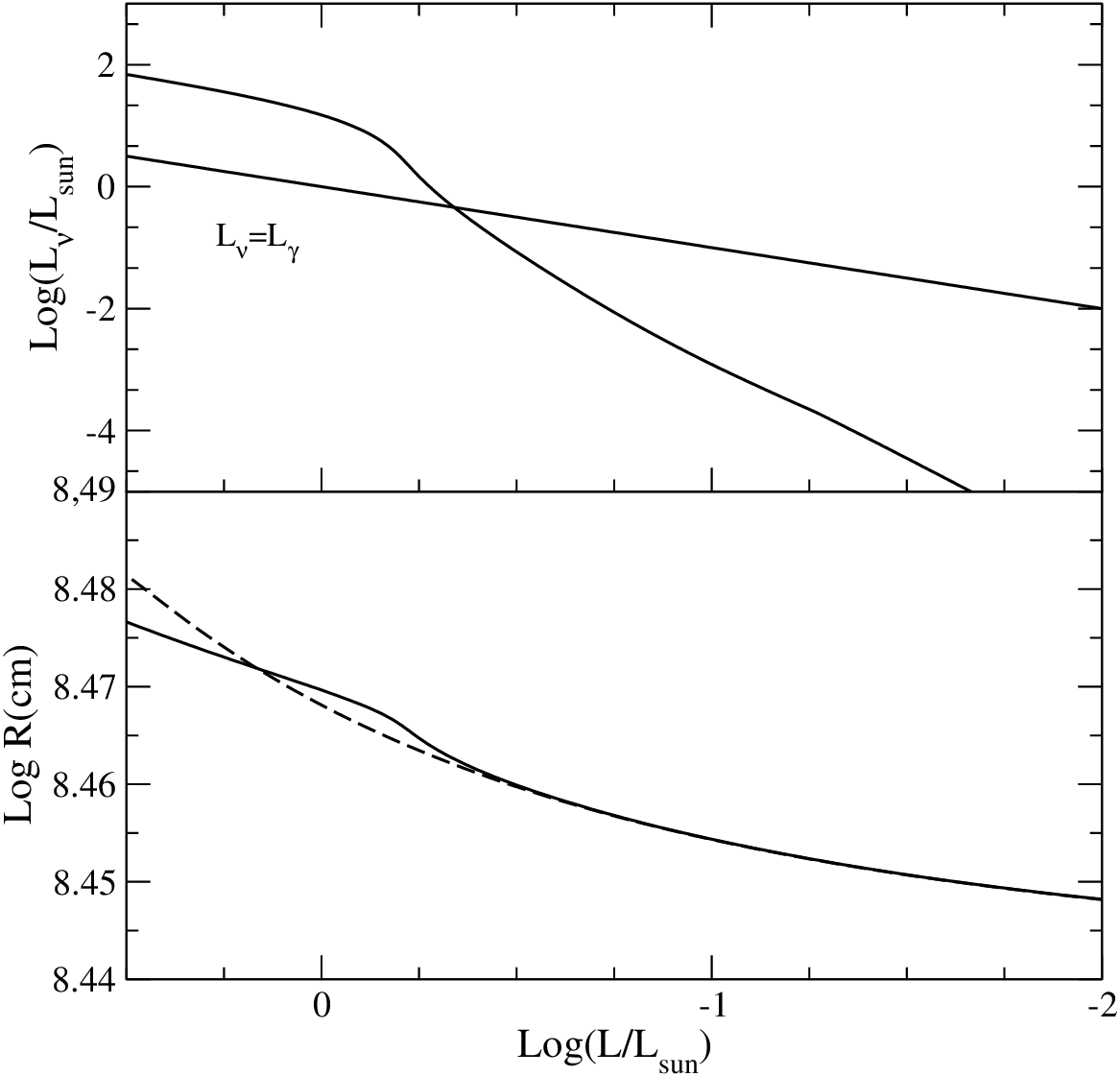}
\caption{Upper panel: Evolution of  the neutrino  emission  luminosity  
         in terms  of the stellar  luminosity for the  $1.28 M_{\sun}$
         white  dwarf sequence.   Bottom  panel: stellar  radius as  a
         function  of  stellar luminosity  for  the $1.28\,  M_{\sun}$
         white dwarf sequence. The dashed line corresponds to the case
         in  which neutrino  losses have  not been  considered  in the
         evolutionary sequence.  Note the change in the stellar radius
         at $\log  L/L_{\sun} \approx  -0.25$ as neutrino  losses fade
         away. }
\label{neurad}
\end{figure}

\begin{figure}[t]
\centering
\includegraphics[clip,width=250pt]{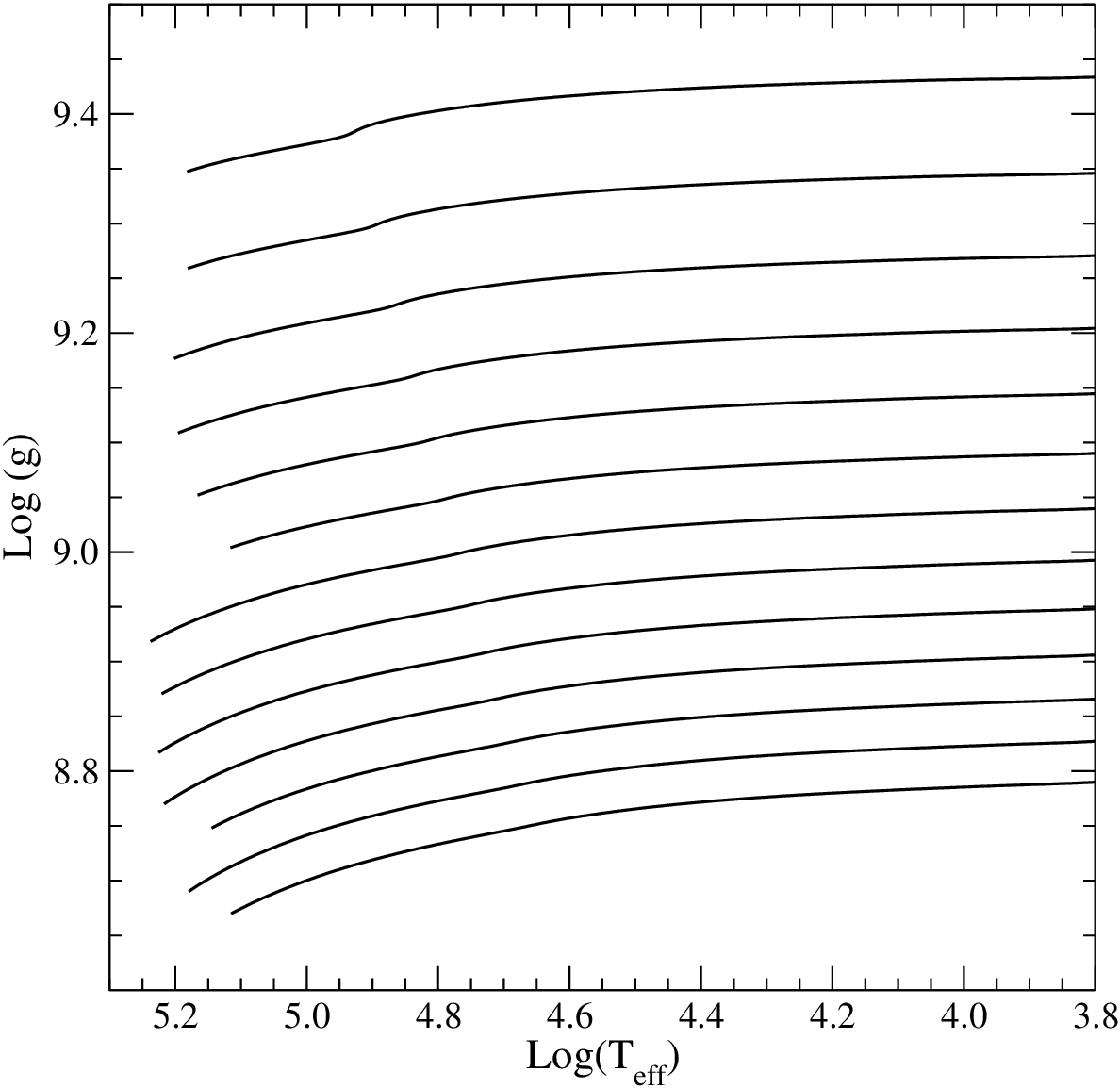}
\caption{Surface  gravity as  a  function of effective temperature for 
         all  of  the   massive  white  dwarf  evolutionary  sequences
         considered  in  this  work.    From  bottom  to  top,  curves
         correspond  to sequences  with  stellar masses  from 1.06  to
         $1.30\, M_{\sun}$  with a mass step of  $0.02\, M_{\sun}$. As
         neutrino   emission  becomes   extinct   at  high   effective
         temperatures, models becomes more compact, as it is reflected
         by the sudden change in the surface gravity, particularly for
         most massive  models. For such sequences  and advanced stages
         of   evolution,    finite-temperature   effects  are   barely
         noticeable.}
\label{gravfull}
\end{figure}

During the first evolutionary stages of the white dwarf models studied
here  ---  those with  $\log  L/L_{\sun}\ga  -0.75$  --- most  of  the
neutrino emission proceeds  via the plasma process, as  it is apparent
from Fig.~\ref{procesos}  (see also Fig.  ~\ref{fig04.eps}).  However,
for  the  most  massive  white  dwarfs, neutrino  losses  through  the
bremsstrahlung process are not negligible whatsoever, being comparable
to or even larger than the  photon luminosity of the white dwarf. This
is  due to the  strong dependence  of the  neutrino emissivity  on the
density and on the average charge of the ions.  Moreover, the neutrino
luminosity far exceeds the photon luminosity during these early stages
of the evolution.

Neutrino losses  constitute a powerful  sink of energy in  hot massive
white dwarfs  and leave their signatures in  the mass-radius relations
of the sequences, as it is apparent from Fig.~\ref{neurad}. Indeed, as
neutrino   emission  gradually   decreases  the   models   undergo  an
appreciable change in their radius.  This effect is clearly noticeable
in the $1.28\,  M_{\sun}$ white dwarf model shown  in the bottom panel
of Fig.~\ref{neurad}, which shows a  marked reduction in the radius of
the  model  as  the  neutrino  emission becomes  extinct  below  $\log
L/L_{\sun} \approx  -0.25$.  Note that  this reduction in  the stellar
radius  is   absent  if  neutrino  losses  are   not  considered  (the
corresponding evolutionary sequence is represented by a dashed line in
the botton panel of  Fig.~\ref{neurad}).  This coupling,  which is
not found for stellar masses smaller than about $1.20\, M_{\sun}$, has
not  been reported  before.  We  stress  that this  behaviour is  only
noticeable  for very  massive  white dwarf  models,  because in  these
models  the maximum  of neutrino  emission occurs  in the  outer, less
degenerate regions of the core. On the contrary, in less massive white
dwarfs, neutrino energy losses are large in the degenerate core of the
white  dwarf. The  upper panel  of Fig.~\ref{neurad}  illustrates the
behaviour of  neutrino luminosity, $L_{\nu}$, as a  function of photon
luminosity.   Here, the various  neutrino processes  are shown  in the
same plot.  Note  that neutrino energy losses in  massive white dwarfs
far  exceed the  photon luminosity.   As the  evolution  proceeds, the
neutrino luminosity  fades away faster than the  photon luminosity and
neutrino cooling has little effect  on the subsequent evolution of the
models.   This behaviour  can  be  explained  with the  help  of
Fig.~\ref{neurad} as  follows. As it  has been already shown,  for the
most massive  ONe white  dwarfs the neutrino  emissivity peaks  in the
outer partially  degenerate layers of  the white dwarf.   Hence, these
layers are cooler than in the case in which neutrino emission has been
disregarded. Consequently,  the radius of  the white dwarf  is smaller
for the  sequence in which  neutrino emission has been  properly taken
into account. Since the peak  of the neutrino emissivity is located at
the edge of the degenerate core,  in a region in which the temperature
gradient is  large, the nearly isothermal core  provides enough energy
to prevent the temperature to  drop very rapidly in these regions.  As
a consequence the rate of change of the stellar radius is smaller than
in the case in which neutrinos are disregarded.  Moreover, since these
layers are  already cooler than those  of the model  in which neutrino
emission is disregarded the rate of change of the radius is smaller as
well. Nevertheless, in these layers neutrino cooling is efficient and,
therefore, the temperature decreases and the density increases.  This,
ultimately  leads to  a  decrease of  the  neutrino luminosity,  which
finally drops below  the photon luminosity.  At this  stage the radius
of the evolutionary sequence in which neutrinos are taken into account
is   indistinguishable  from   that  in   which  neutrinos   has  been
disregarded. 

The  main result  of our  evolutionary calculations  is  summarized in
Fig.~\ref{gravfull}, which shows the  evolution of the surface gravity
(in cgs  units) for all our  sequences as a function  of the effective
temperature.  As it is well  known, for a given effective temperature,
more  massive models  are characterized  by smaller  radii  and larger
gravities.  As cooling gradually proceeds, the surface gravity becomes
larger, eventually reaching an  almost constant value corresponding to
that of  the zero-temperature configuration, as it  should be expected
for strongly degenerate  configurations.  Note that finite-temperature
effects are  appreciable only at  high effective temperatures,  in the
early evolutionary phases.

\begin{figure}[t]
\centering
\includegraphics[clip,width=250pt]{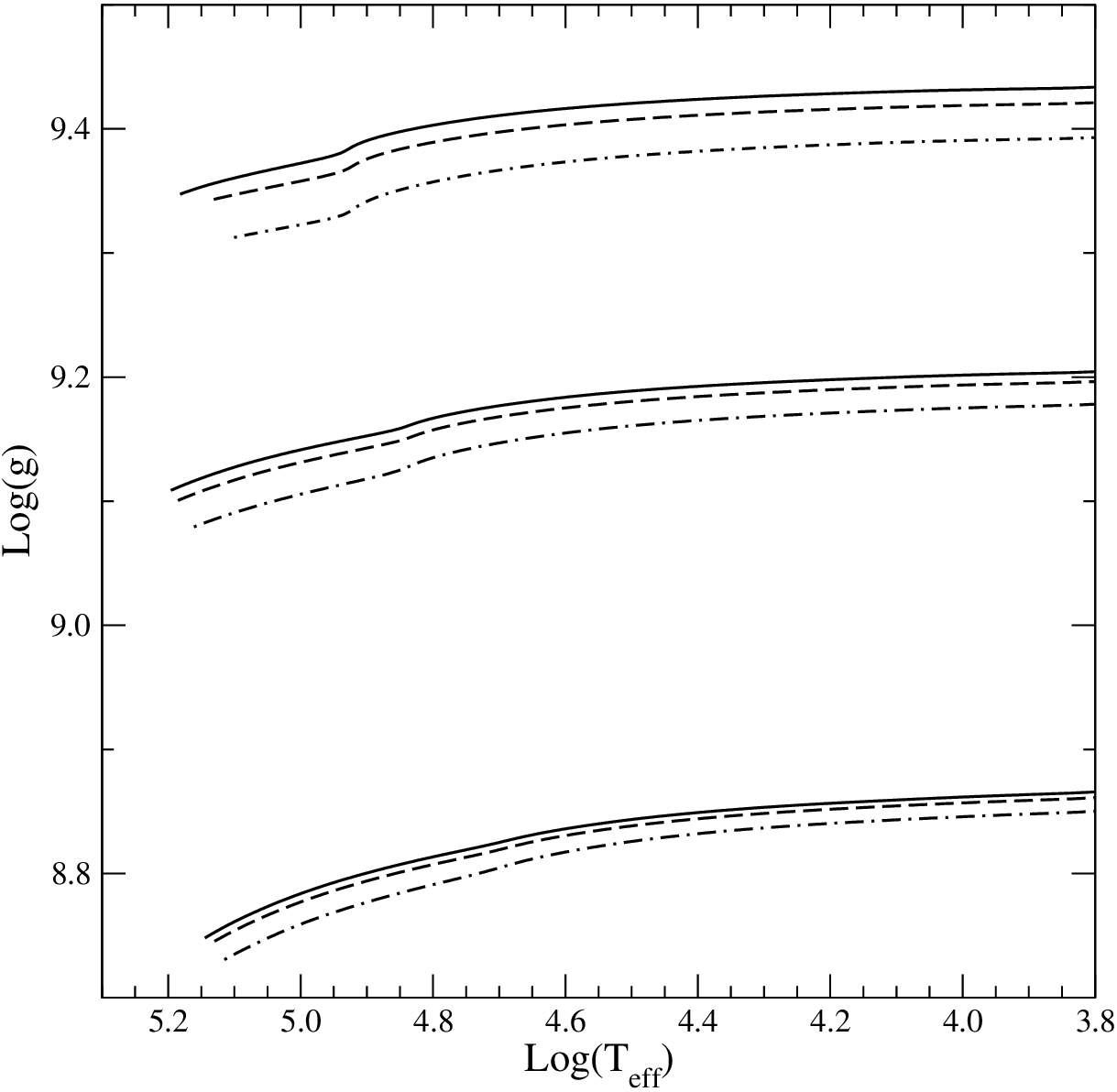}
\caption{Surface  gravity as  a  function of effective temperature for 
         the 1.30,  1.24 and  $1.10\, M_{\sun}$ white  dwarf sequences
         with  various core chemical  compositions. Solid,  dashed and
         dot-dashed   lines   correspond   to  core   compositions  of
         oxygen-neon,  oxygen  and  carbon,  respectively.   Note  the
         effect of  the core chemical composition  on surface gravity,
         particulary for very massive white dwarfs.}
\label{compo}
\end{figure}

Because  of the  presence of  abundant $^{20}$Ne  in the  core  of our
massive white  dwarf models, gravities are larger  than those obtained
when  a CO  core  is adopted.   This  is a  result  of the  non-ideal,
corrective terms  of the equation of  state. In fact,  the presence of
$^{20}$Ne yields  a much  stronger interacting plasma  than that  of a
regular CO mixture adopted in  the previous works (Wood 1995). This is
borne out  by Fig.~\ref{compo} which illustrates the  evolution of the
surface gravity for selected massive white dwarf models with different
core chemical composition.  Specifically, solid, dashed and dot-dashed
lines  correspond  to core  compositions  of  oxygen-neon, oxygen  and
carbon, respectively.  As expected,  for a given stellar mass, surface
gravities are larger the higher  the atomic number $Z$ of the chemical
constituent.   Interestingly  enough, very  massive  white dwarfs  are
rather sensitive to the  adopted core chemical composition. Such white
dwarfs are characterized by densities  high enough for electrons to be
extremely  relativistic throughout  most of  their  interiors.  Hence,
Coulomb interactions grow in importance as the stellar mass increases.
Finally, in  table 1  we list the  surface gravity values  expected at
\teff= 20\,000K for some selected white dwarf sequences with different
core chemical composition.

\begin{table}
\caption{Surface gravities for white dwarf  models with different core 
chemical composition at \teff= 20\,000K. }
\renewcommand{\arraystretch}{1.3}
\begin{tabular}{p{1.5cm}p{1.5cm}p{1.5cm}}
\hline
\hline
$ M_*/M_{\sun}$ & $X_{\rm core}$ & $\log(g)$\\
\hline
1.10  & ONe & 8.853  \\
      & O   & 8.848  \\
      & C   & 8.837  \\ 
1.24  & ONe & 9.195  \\
      & O   & 9.187  \\
      & C   & 9.168  \\
1.30  & ONe & 9.426  \\
      & O   & 9.414  \\
      & C   & 9.385  \\
\hline
\hline
\end{tabular}
\end{table}

\section{Conclusions}

We have computed reliable  mass-radius relations --- or, equivalently,
$\log  g$/$T_{\rm eff}$ relations  --- for  massive white  dwarfs with
degenerate ONe cores.  Our  calculations encompass masses ranging from
1.06 to $1.30\, M_{\sun}$, the  expected range of masses for which ONe
white dwarfs  should presumably  exist. To the  best of  our knowledge
this is the first attemtpt to compute such a relation with a realistic
equation of  state --- which includes the  non-ideal, corrective terms
and the full temperature dependence --- and reliable chemical profiles
for  the  degenerate  interior  reflecting the  previous  evolutionary
history  of  the  progenitors  of  such  white  dwarfs.   The  results
presented here  will undoubtedly be  helpful in the  interpretation of
recent  observations  of white  dwarf  stars  with  very high  surface
gravities (Dahn  et al.  2004; Madej  et al.  2004;  Nalezyty \& Madej
2004), which  up to now  relies on prior evolutionary  sequences which
were computed  assuming CO  cores.  Moreover, we  have shown  that the
mass-radius  relation  shows  a  marked  dependence  on  the  neutrino
luminosity, in such a way that  for the most massive stars within this
mass range a sudden decrease  in the mass-radius relationship shows up
shortly  before  the  neutrino   luminosity  drops  below  the  photon
luminosity.

Finally, we  have prepared detailed tabulations of  radius and gravity
for all our massive white  dwarf sequences, which are available at our
web site: {\tt http://www.fcaglp.unlp.edu.ar/evolgroup/}.

\begin{acknowledgements}
This   research  was   partially   supported  by   the  Instituto   de
Astrof\'{\i}sica La Plata, by  the MCYT grant AYA04094--C03-01 and 02,
by  the  European  Union  FEDER  funds,  and  by  the  CIRIT.   L.G.A.
acknowledges the Spanish MCYT for a Ram\'on y Cajal Fellowship.
\end{acknowledgements}

\end{document}